\documentclass[pra,twocolumn,showpacs,preprintnumbers,amsmath,amssymb,nofootinbib]{revtex4}
\usepackage{slashed}
\usepackage{graphicx}
\usepackage{psfrag}
\usepackage{hyperref}

\begin{document}

\title{Modified Fermi-sphere, pairing gap and critical temperature for the BCS-BEC crossover} \date{\today} \author{S. Floerchinger${}^1$, M. M. Scherer${}^2$ and C. Wetterich${}^1$}


\affiliation{\mbox{\it ${}^1$Institut f{\"u}r Theoretische Physik, Universit{\"a}t Heidelberg,
Philosophenweg 16, D-69120 Heidelberg, Germany} 
\mbox{\it ${}^2$Theoretisch-Physikalisches Institut, Friedrich-Schiller-Universit{\"a}t Jena,
Max-Wien-Platz 1, D-07743 Jena, Germany}
\mbox{\it E-mail: {s.floerchinger@thphys.uni-heidelberg.de, michael.scherer@uni-jena.de, c.wetterich@thphys.uni-heidelberg.de}}
}
\begin{abstract} 
We investigate the phase diagram of two-component fermions in the BCS-BEC crossover. Using functional renormalization group equations we calculate the effect of quantum fluctuations on the fermionic self-energy parametrized by a wavefunction renormalization, an effective Fermi radius and the gap. This allows us to follow the modifications of the Fermi surface and the dispersion relation for fermionic excitations throughout the whole crossover region. We also determine the critical temperature of the second order phase transition to superfluidity. Our results are in agreement with BCS theory including Gorkov's correction for small negative scattering length $a$ and with an interacting Bose gas for small positive $a$. At the unitarity point the result for the gap at zero temperature agrees well with Quantum-Monte-Carlo simulations while the critical temperature differs.
\end{abstract}

\maketitle

\section{Introduction}

In the vicinity of a Feshbach resonance cold fermionic atom gases show a continuous crossover from BCS to BEC condensation \cite{p5ALeggett80}. This qualitative picture has been confirmed in recent experiments \cite{Exp}. As the system can serve as a testing ground for non-perturbative methods a quantitatively precise theoretical understanding of this phenomenon is the aim of various theoretical approaches. At and near the resonance quantum Monte-Carlo (QMC) results have been obtained
\cite{MonteCarlo,MonteCarloBuro,MonteCarloBulgac,MonteCarloBulgac2,MonteCarloAkki}. The complete phase diagram has been accessed by functional field-theoretical techniques, in particular by $\epsilon$-expansion
\cite{Nussinov, Nishida:2006br, Nishida:2006eu, EpsilonExpansion},
$1/N$-expansion \cite{Nikolic:2007zz, AbukiBrauner}, $t$-matrix approaches
\cite{TMatrix,Perali:2004zz}, Dyson-Schwinger equations \cite{Diehl:2005ae, Diener08},
2-Particle Irreducible methods \cite{Haussmann:2007zz}, and nonperturbative
renormalization-group flow equations \cite{Birse05, Diehl:2007th, Diehl:2007XXX, Gubbels:2008zz, FSDW08, Bartosch:2009zr}.

The techniques and results from renormalization-group flow equations for cold atoms have been gathered and reviewed in \cite{Diehl:2009ma}. There, a basic truncation is investigated and extended to include important quantitative effects, such as particle-hole fluctuations. In the present work we aim at a systematic completion of the truncation scheme and consider further important, however subleading terms in the flowing action. In the fermionic sector we include a renormalization term adding  to the chemical potential and the fermionic wave function renormalization. This accounts for changes of the Fermi surface due to fluctuation effects. Similar parametrizations of the fermionic self-energy have been studied in refs.\ \cite{Gubbels:2008zz, Bartosch:2009zr, Strack08}. In ref.\ \cite{Spectraldensities} the spectral function of ultracold fermions in the BCS-BEC crossover has been calculated from a ladder approximation to the Luttinger-Ward functional. 

As an additional parameter, we also include an atom-dimer interaction term. These extensions of the truncation improve our numerical results in the perturbative domains of the crossover to match the results from well-known other field-theoretical methods. We obtain considerable quantitative precision on the BCS and BEC sides of the resonance. Also in the strongly interacting regime of the Feshbach resonance we obtain a good agreement with data from QMC simulations for the gap and the chemical potential at zero temperature. The effect of the additional couplings on the critical temperature at the unitarity point, however, is rather modest and no agreement with the widely accepted QMC simulations is found.

For a microscopic model of the BCS-BEC crossover we employ a two-component Grassmann field $\psi=(\psi_1,\psi_2)$, describing fermions in two hyperfine states.  As bosonic degrees of freedom we introduce a complex scalar field $\phi$, which can describe molecules, Cooper pairs or simply an auxiliary field in different regimes of the crossover. The resulting two-channel model can describe both narrow and broad Feshbach resonances in a unified setting. The microscopic action at the ultraviolet scale $\Lambda$ reads explicitely
\begin{eqnarray}
\nonumber
S[\psi, \phi] & = & \int_0^{1/T} d\tau \int d^3x{\Big \{}\psi^\dagger(\partial_\tau-\Delta-\mu)\psi\\
\nonumber
& & +\phi^*(\partial_\tau-\frac{1}{2}\Delta-2\mu+ \nu_\Lambda)\phi\\
& & - h_\Lambda(\phi^*\psi_1\psi_2+h.c.){\Big \}}\,,
\label{eqMicroscopicAction}
\end{eqnarray}
where we choose nonrelativistic natural units with $\hbar=k_B=2M=1$, with $M$ the mass of the atoms. and $\Delta$ denotes the Laplace operator.
We use the Matsubara formalism to describe the system in thermal equilibrium, so the fields depend on the position variable $\vec{x}$ and  on the imaginary time variable $\tau$, which parameterizes a torus with circumference $1/T$. The variable $\mu$ is the chemical potential. We introduce a Yukawa or Feshbach interaction $h$ which couples the fermionic to the bosonic fields and directly relates to the width of the Feshbach resonance. The parameter $\nu$ depends on the magnetic field and accounts for the detuning from the Feshbach resonance. The microscopic values $h_\Lambda$, $\nu_\Lambda$ can be fixed by the properties of two body scattering in vacuum. This is discussed in detail in \cite{Diehl:2007th,Diehl:2009ma}. 

Here we will discuss the limit of broad Feshbach resonances, which corresponds to $h_\Lambda\to\infty$. Then the microscopic interaction between fermions becomes pointlike, with strength $-h_\Lambda^2/\nu_\Lambda$. The system shows a far reaching universality at a broad Feshbach resonance such that macroscopic quantities are independent of the microscopic details. In our case the relevant parameter is the two-body scattering length $a$ or, at finite density, the concentration $c=ak_F$, where the formal Fermi momentum is related to the density by $k_F=(3\pi^2 n)^{1/3}$. At nonzero temperature we have an additional parameter given by $T/T_F$, where $T_F=E_F=k_F^2$  is the Fermi temperature. 

The paper is organized as follows: In Sect.\ \ref{sec:floweq} we introduce the exact functional flow equation and specify the truncation as well as our regulator scheme. Section \ref{sec:bosonization} discusses the treatment of particle-hole fluctuations by a scale-dependent Hubbard-Stratonovic transformation using a recently derived exact flow equation. We present our results for the fermionic self-energy, the single-particle gap and the critical temperature throughout the crossover in Sect.\ \ref{sec:results} and finally draw some conclusions in Sect. \ref{sec:conc}.

\section{Flow equation \& Truncation}
\label{sec:floweq}

The functional renormalization group connects the microphysics, cf. equation \eqref{eqMicroscopicAction}, to macrophysics, i.e. to observable thermodynamics, by means of a non-perturbative flow equation. The scale dependence of the flowing or average action is given by the exact functional flow equation \cite{Wetterich:1992yh}

\begin{equation}\label{eq:flowequwett}
\partial_k \Gamma_k[\chi]=\frac{1}{2}\mathrm{STr}\left[ \left(\Gamma_k^{(2)}+R_k \right)^{-1} \partial_k R_k\right] \,.
\end{equation}
Here, the STr operation involves an integration over momenta as well as a summation over internal indices, with an appropriate minus sign for the fermions. The infrared cutoff function $R_k(q)$ has the properties $R_k(q) \rightarrow \infty$ for $k \rightarrow \Lambda$, $R_k(q) \approx k^2$ for $\frac{|q|}{k} \rightarrow 0$ and $R_k(q) \rightarrow 0$ for $\frac{k}{|q|} \rightarrow 0$. We employ the second functional derivative of $\Gamma_k$

\begin{equation}
\left( \Gamma_k^{(2)}[\chi]\right)_{ij}(p_1,p_2)=\frac{\overrightarrow\delta}{\delta\chi_{i}(-p_1)}\Gamma_k[\chi]\frac{\overleftarrow\delta}{\delta\chi_{j}(p_2)}\,.
\end{equation}
Equation (\ref{eq:flowequwett}) is an exact differential equation for a functional. This corresponds to a system of infinitely many coupled differential equations for running couplings. Finding exact solutions to equation (\ref{eq:flowequwett}) for non-trivial theories is possible in practice, but one can use truncations in the space of possible functionals to find approximate solutions. Such approximations do not have to rely on the existence of a small expansion parameter like the interaction strength and they are therefore of a non-perturbative nature. For reviews of the functional renormalization group method see \cite{Berges:2000ew,Pawlowski:2005xe}.


In this work, we solve the flow equation \eqref{eq:flowequwett} by using a truncation which is a completion of the truncation scheme that was discussed in detail in \cite{FSDW08, Diehl:2009ma}. More explicitly,
\begin{eqnarray}
\nonumber
\Gamma_k[\chi] & =  & \int_0^{1/T}d\tau \int d^3x {\bigg \{} \bar\psi^\dagger Z_\psi(\partial_\tau -\Delta) \bar\psi +\bar m_\psi^2\bar\psi^\dagger\bar\psi\\
\nonumber
& + & \bar{\phi}^*(\bar Z_\phi \partial_\tau-\frac{1}{2}A_\phi \Delta)\bar\phi + \bar U(\bar \rho,\mu) \\
& - & \bar h (\bar \phi^* \bar\psi_1\bar\psi_2 + \bar \phi \bar\psi_2^\ast \bar\psi_1^\ast )+\bar\lambda_{\phi\psi}\bar\phi^\ast\bar\phi\bar\psi^\dagger\bar\psi {\bigg \} }.
\label{eq:baretruncation}
\end{eqnarray}
The effective potential $\bar U(\bar \rho,\mu)$ contains no derivatives and is a function of $\bar{\rho}=\bar{\phi}^*\bar{\phi}$ and $\mu$. Besides the couplings parameterizing $\bar U$ (see below) our truncation contains six further $k$-dependent (``running'') couplings $Z_\psi$, $\bar m_\psi^2$, $A_\phi$, $\bar Z_\phi$, $\bar h$, and $\bar \lambda_{\phi\psi}$. In addition to the parameters used in our previous investigations \cite{FSDW08, Diehl:2009ma}, our truncation includes now the $k$-dependent parameters $\bar m_\psi^2$ and $Z_\psi$ by which we take into account fluctuation effects on the self-energy of the fermionic quasiparticles. The inclusion of the coupling constant $\bar \lambda_{\phi\psi}$ closes the truncation on the level of interaction terms quartic in the fields (see also \cite{Krippa:2009vu}). The coupling $\bar \lambda_{\phi\psi}$ plays an important role for the three-body physics \cite{Diehl:2007xz, MFSW09} and is expected to lead to quantitative modifications for the many-body problem. In contrast to \cite{Diehl:2007xz,MFSW09} we do not resolve the momentum dependence of $\bar\lambda_{\phi\psi}$ in this work.

Another quantitative contribution to the many-body physics of the BCS-BEC crossover is given by particle-hole fluctuations. We will include them by the technique of scale dependent bosonization. A brief discussion of this is given in the next section. For a more detailed discussion we refer to a previous paper \cite{FSDW08}.

In terms of renormalized fields $\phi=A_\phi^{1/2}\bar\phi$, $\rho=A_\phi \bar \rho$, $\psi=Z_\psi^{1/2}\bar\psi$, renormalized couplings $Z_\phi=\bar Z_\phi / A_\phi$, $h=\bar h/(A_\phi^{1/2} Z_\psi)$, $\lambda_{\phi\psi}=\bar \lambda_{\phi\psi}/(A_\phi Z_\psi)$, $m_\psi^2=\bar m_\psi^2/Z_\psi$ and effective potential $U(\rho,\mu)=\bar U(\bar \rho, \mu)$, Eq. \eqref{eq:baretruncation} reads

\begin{eqnarray}
\nonumber
\Gamma_k[\chi] & =  & \int_0^{1/T}d\tau \int d^3x {\bigg \{} \psi^\dagger (\partial_\tau -\Delta+m_\psi^2) \psi \\
\nonumber
& + & {\phi}^*(Z_\phi \partial_\tau-\frac{1}{2} \Delta)\phi + U(\rho,\mu) \\
& - & h ( \phi^* \psi_1\psi_2 + \phi \psi_2^\ast \psi_1^\ast )+\lambda_{\phi\psi}\phi^\ast\phi\psi^\dagger\psi {\bigg \} }.
\label{eq:truncation}
\end{eqnarray}
For the effective potential, we use an expansion around the $k$-dependent location of the minimum $\rho_0(k)$ and the $k$-independent value of the chemical potential $\mu_0$ that corresponds to the physical particle number density $n$. We determine $\rho_0(k)$ and $\mu_0$ by 
\begin{eqnarray}
\nonumber
(\partial_\rho U)(\rho_0(k),\mu_0)=0 &&\text{for all }k\\
-(\partial_\mu U) (\rho_0,\mu_0)=n && \text{at }k=0.  
\end{eqnarray}
More explicitly, we employ a truncation for $U(\rho,\mu)$ of the form
\begin{eqnarray}
\nonumber
U(\rho,\mu) &=& U(\rho_0,\mu_0)-n_k (\mu-\mu_0)\\
&& +(m^2+\alpha(\mu-\mu_0)) (\rho-\rho_0)\nonumber\\
&& +\frac{1}{2}\lambda (\rho-\rho_0)^2.
\end{eqnarray}
This expansion is discussed in detail in refs.\ \cite{FloerchingerWetterich, FWThermod}.
In the symmetric or normal gas regime, we have $\rho_0=0$, while in the regime with spontaneous symmetry breaking, we have $m^2=0$. The atom density $n=-\partial U/\partial \mu$ corresponds to $n_k$ in the limit $k\to 0$. The system is superfluid if $\rho_0(k\to 0)>0$. For $\rho_0>0$ the term $\sim \lambda_{\phi\psi}$ leads to a further modification of the Fermi surface, besides the gap $\sqrt{h^2 \rho_0}$. 

In total, we have eight running couplings $m^2(k)$, $m_\psi^2(k)$, $\lambda(k)$, $\alpha(k)$, $n_k$, $Z_\phi(k)$, $h(k)$ and $\lambda_{\phi\psi}(k)$. (In the phase with spontaneous symmetry breaking $m^2$ is replaced by $\rho_0$.)  In addition, we need the anomalous dimensions $\eta_\phi=-k\partial_k \text{ln} A_\phi$ and $\eta_\psi=-k\partial_k \text{ln} Z_\psi$. We project the flow of the average action $\Gamma_k$  on the flow of these couplings by taking appropriate (functional) derivatives on both sides of Eq. \eqref{eq:flowequwett}.  We thereby obtain a set of coupled nonlinear differential equations which can be solved numerically. 

At the microscopic scale $k=\Lambda$ the initial values of our couplings are determined from Eq.\ \eqref{eqMicroscopicAction}. This gives $m_\psi^2(\Lambda)=-\mu$, $m^2(\Lambda)=\nu_\Lambda-2\mu$, $\rho_{0,\Lambda}=0$, $\lambda(\Lambda)=0$, $Z_\phi(\Lambda)=1$, $h(\Lambda)=h_\Lambda$, $\alpha(\Lambda)=-2$ and $n_\Lambda=3\pi^2 \mu \theta(\mu)$. The initial values $\nu_\Lambda$ and $h_\Lambda$ can be connected to the two particle scattering in vacuum close to a Feshbach resonance. For this purpose one follows the flow of $m^2(k)$ and $h(k)$ in vacuum, i.e. $\mu=T=n=0$ and extracts the renormalized parameters $m^2=m^2(k=0)$, $h=h(k=0)$. These are connected to the scattering length by 
\begin{equation}
a=-\frac{h^2(k=0)}{8\pi\, m^2(k=0)}.
\end{equation}
For details we refer to ref.\ \cite{Diehl:2009ma}.

The infrared cutoff we use is purely space-like and reads in terms of the bare fields
\begin{eqnarray}
\nonumber
\Delta S_k &=& \int_p {\bigg \{} \bar\psi^\dagger(p)Z_\psi\big[\text{sign}(\vec p^2-\tilde \mu_k)k^2\\
\nonumber
&-& (\vec p^2-\tilde \mu_k)\big)]\theta\left(k^2-|\vec p^2-\tilde \mu_k|\right)\bar\psi(p)\\
\nonumber
&+& \bar \phi^*(p)A_\phi \left[k^2-\vec p^2/2\right]\theta\left(k^2-\vec p^2/2\right) \bar \phi(p) {\bigg \}},\\
\tilde \mu_k &=& - m_\psi^2-\lambda_{\phi\psi}\rho_0.
\label{eq:cutoff}
\end{eqnarray}
For the fermions it regularizes fluctuations around the running Fermi surface, while for the bosons fluctuations with small momenta are suppressed. The choice of $\Delta S_k$ in Eq. \eqref{eq:cutoff} is an optimized choice in the sense of \cite{Litim:2000ci,Litim:2001up,Pawlowski:2005xe}.

\section{Scale dependent bosonisation}
\label{sec:bosonization}

\begin{figure}[t!]
\centering
\includegraphics[width=0.3\textwidth]{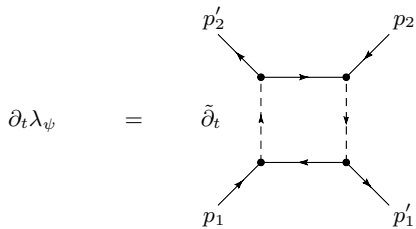}
\caption{Box diagram for the flow of the four-fermion interaction.}
\label{fig:boxes}
\end{figure}

In the microscopic model in Eq.\ \eqref{eqMicroscopicAction} a Hubbard-Stratonovich transformation relates the model for fermions with a local interaction to our Yukawa-like theory. The fermion-fermion interaction has been replaced by a tree-level process that involves the composite boson $\phi$. In the limit of vanishing temperature and density (few-body physics) this leads to a consistent treatment since an explicit fermion-fermion interaction term in the effective action vanishes on all scales. At nonzero density the situation is different, however. Particle-hole fluctuations generate a term of the form
\begin{equation}
\lambda_\psi\; \int_0^{1/T}d\tau d^3 x\; \psi^*_1 \psi_1 \psi^*_2 \psi_2
\end{equation}
in the effective action, although it vanishes in the microscopic action, cf. Fig. \ref{fig:boxes}. In principle one could simply take $\lambda_\psi$ as an additional coupling into account. However, it is much more elegant to use a scale-dependent Hubbard-Stratonovich transformation \cite{Gies:2001nw,FWCompositeoperators} which absorbes $\lambda_\psi$ into the Yukawa-type interaction with the bosons at every scale $k$. By construction, there is then no self interaction between the fermionic quasiparticles. The general procedure of ``partial bosonization'' is discussed in detail in \cite{FSDW08}. In the present paper we employ the exact flow equation derived in \cite{FWCompositeoperators}. Compared to the scheme used in \cite{FSDW08} the flow equations are modified slightly. Following the steps in \cite{FSDW08, FWCompositeoperators} we find for the renormalized coupling $m^2$ in the symmetric regime an additional term reflecting the absorption of $\lambda_\psi$ into the fermionic interaction induced by boson exchange,
\begin{equation}
\partial_t m^2 = \partial_t m^2{\big |}_\text{HS} + \frac{m^4}{h^2} \partial_t \lambda_\psi {\big |}_\text{HS}.
\end{equation}
Here $\partial_t m^2{\big |}_\text{HS}$ and $\partial_t \lambda_\psi {\big |}_\text{HS}$ denote the flow equations when the Hubbard-Stratonovich transformation is kept fixed. The flow equation of all other couplings are the same as with fixed Hubbard-Stratonovich transformation. In the regime with spontaneous symmetry breaking we use
\begin{eqnarray}
\nonumber
\partial_t h &=& \partial_t h {\big |}_\text{HS} + \frac{\lambda \rho_0}{h} \partial_t \lambda_\psi {\big |}_\text{HS},\\
\nonumber
\partial_t \rho_0 &=& \partial_t \rho_0{\big |}_\text{HS} - 2 \frac{\lambda \rho_0^2}{h^2} \partial_t \lambda_\psi{\big |}_\text{HS},\\
\partial_t \lambda &=& \partial_t \lambda{\big |}_\text{HS} + 2 \frac{\lambda^2 \rho_0}{h^2} \partial_t \lambda_\psi{\big |}_\text{HS}.
\end{eqnarray}

\section{Results}
\label{sec:results}

The extension of the truncation by including the running couplings $Z_\psi, m_\psi^2$ and $\lambda_{\phi\psi}$ and the different bosonisation scheme as compared to \cite{FSDW08} help to improve the numerical precision of our results considerably all over the crossover, however especially in the limits of weak interactions. The effect of including the $\lambda_{\phi\psi}$-vertex can most prominently be observed in the result for the vacuum dimer-dimer scattering length $a_M/a$. In this truncation we find $a_M/a=0.59$, which is in very good agreement with the well-known result from solving the Schr\"odinger equation, $a_M/a=0.60$ \cite{Petrov04}. Our result is somewhat surprising since no momentum dependence of $\lambda_{\phi\psi}$ has been taken into account while this turns out to be quite important for the atom-dimer scattering length \cite{Diehl:2007xz}.

\begin{figure}[t!]
\centering
\includegraphics[scale=0.98]{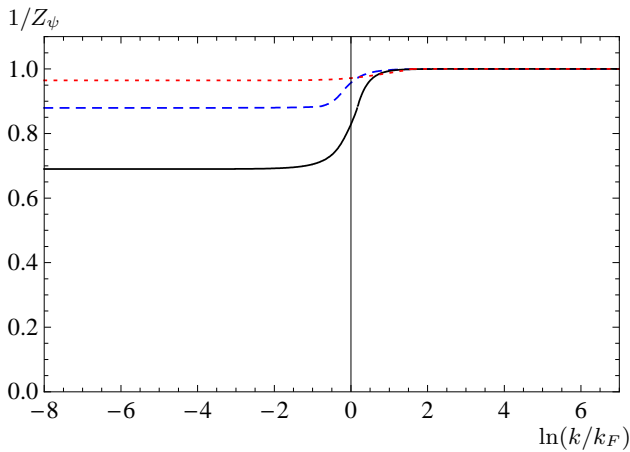}
\caption{Flow of the inverse fermionic wave function renormalization $1/Z_\psi$ at $T=0$. The solid line is at the unitarity point ($c^{-1}=0$). The dashed line corresponds to the BCS side at $c^{-1}=-1$ and the short dashed line to the BEC side at $c^{-1}=1$.}
\label{fig:zpsiflow}
\end{figure}

\begin{figure}[b!]
\centering
\includegraphics[scale=0.99]{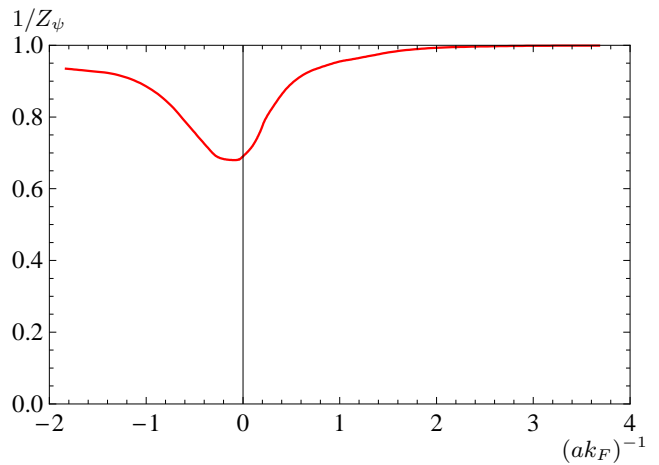}
\caption{Inverse fermionic wave function renormalization $1/Z_\psi$ at the macroscopic scale $k=0$ as a function of the crossover parameter $(ak_F)^{-1}$.}
\label{fig:fermwave}
\end{figure}

A study of the flow of the wave function renormalization $Z_\psi$ shows, that its effect is strongest in the strongly interacting regime, when $|c^{-1}|<1$, cf. Fig. \ref{fig:zpsiflow}. Away from the unitarity point $Z_\psi$ has only a weak scale dependence and stays close to its initial value $Z_\psi(k=\Lambda)=1$, see Fig. \ref{fig:fermwave}.

\subsection{Fermi sphere}
\begin{figure}[t!]
\centering
\includegraphics[scale=0.98]{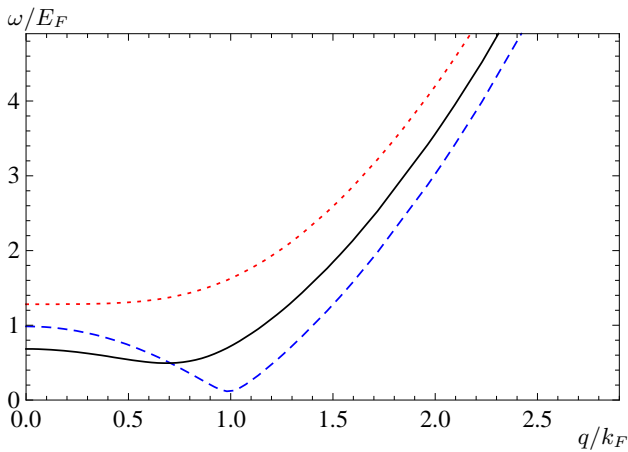}
\caption{Positive branch of the dispersion relation $\omega(q)$ in units of the Fermi energy as given in Eq.\ \eqref{eq:disprel} for $c^{-1}=-1$ (dashed), $c^{-1}=0$ (solid) and $c^{-1}=1$ (short dashed). The gap and the Fermi radius are evaluated at the macroscopic scale $k=0$.}
\label{fig:disprel}
\end{figure}

The flowing coupling constants $Z_\psi$ and $m_\psi^2$ that parametrize the effect of fluctuations on the fermionic self-energy as well as the coupling $\lambda_{\phi\psi}$ have an interesting effect on the dispersion relation of the fermions. In the regime with spontaneously broken $U(1)$ symmetry, the renormalized propagator of the fermionic field reads after analytic continuation to real frequencies $\omega$
\begin{equation}
G_\psi^{-1} = \begin{pmatrix} -h \phi_0 \epsilon,\quad-\omega -(\vec q^2+m_\psi^2+\lambda_{\phi\psi} \rho_0) \\ -\omega + (\vec q^2+m_\psi^2+\lambda_{\phi\psi} \rho_0),\qquad\;\; h \phi_0 \epsilon \end{pmatrix}.
\end{equation}
The dispersion relation follows from $\text{det}\,G_\psi^{-1}=0$ as 

\begin{equation}\label{eq:disprel}
\omega = \pm \sqrt{\Delta^2+(\vec q^2-r_F^2)^2}
\end{equation}
where $\Delta=h\sqrt{\rho_0}$ is the gap and $r_F=\sqrt{-m_\psi^2-\lambda_{\phi\psi}\rho_0}$ is the effective radius of the Fermi sphere. 
\begin{figure}[t!]
\centering
\includegraphics[scale=0.99]{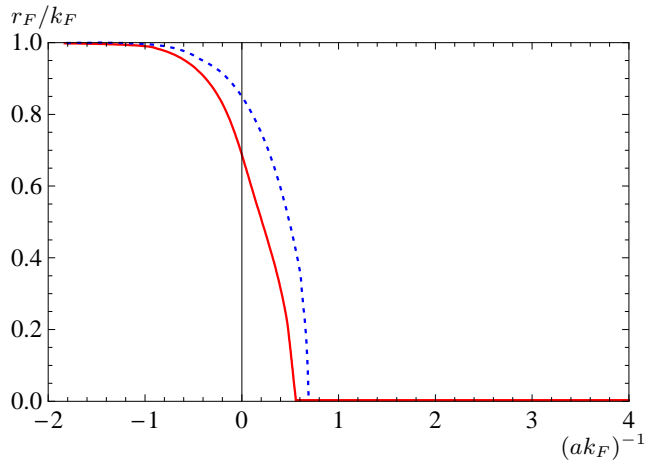}
\caption{Effective Fermi radius $r_F/k_F$ as a function of the crossover parameter $(ak_F)^{-1}$ for vanishing temperature (solid line). We find that the Fermi sphere vanishes approximately at the point with  $(ak_F)^{-1}\approx 0.6$. For comparison we also give the effective Fermi radius in an approximation, where we omitted the contribution of the atom-dimer vertex $\lambda_{\phi\psi}$ (dotted line).}
\label{fig:fermiradius}
\end{figure}
For small negative scattering length on the BCS side of the crossover the renormalization effects on $r_F$ remain small. It takes the classical value $r_F=\sqrt{\mu}=k_F$ where $k_F=(3\pi^2 n)^{1/3}$. For large negative scattering length the Fermi sphere gets smaller and $r_F$ finally vanishes at a point with $(ak_F)^{-1}\approx 0.6$. For small and positive scattering length (BEC side), the Fermi surface has disappeared and the fermions are gapped even for $\Delta\to0$ by a positive value of $m_\psi^2+\lambda_{\phi\psi} \rho_0$. We plot our result for the Fermi radius in units of the Fermi momentum $k_F=(3\pi^2 n)^{1/3}$ in Fig. \ref{fig:fermiradius} and the dispersion relation in Fig. \ref{fig:disprel}. 

In order to show explicitely the effect of the atom-dimer vertex $\lambda_{\phi\psi}$ on the radius of the Fermi sphere we also plot $r_F$ as $\lambda_{\phi\psi}$ is omitted in the truncation (Fig. \ref{fig:fermiradius}, dotted line). We find that this changes the estimate for $r_F/k_F$ quantitatively for crossover parameters $(ak_F)^{-1} > -1$. We conlude that the atom-dimer vertex affects not only the physics on the BEC side as reflected for example in the ratio $a_M/a$, but also has an effect close to unitarity.

\subsection{Single-particle gap at $T=0$}
To reduce the numerical effort we have omitted the effect of particle-hole fluctuations for the study of the effects discussed so far. In the following all results are given inluding particle-hole fluctuations, however.

\begin{figure}[t!]
\centering
\includegraphics[scale=0.99]{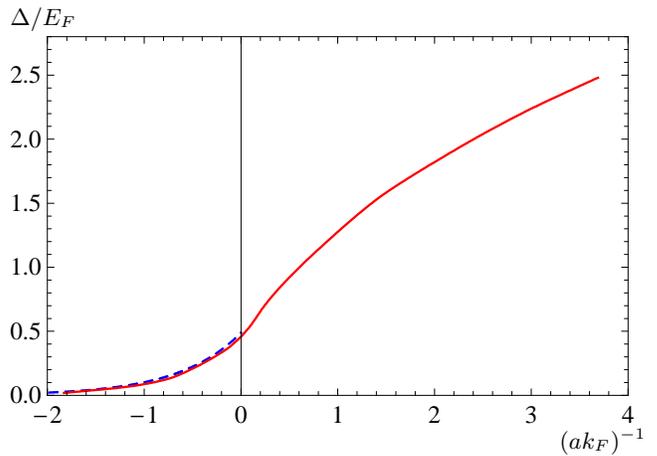}
\caption{Gap in units of the Fermi energy $\Delta/E_F$ as a function of $(ak_F)^{-1}$ (solid line). For comparison, we also plot the result found by Gorkov and Melik-Bakhudarov (left dahed) and extrapolate this to the unitarity point $(ak_F)^{-1}=0$, where $\Delta_{\mathrm{GMB}}/E_F=0.49$.}
\label{fig:gap}
\end{figure}

On the BCS side for small negative values of $ak_F$ the single particle gap at zero temperature can be calculated from perturbation theory. The result first found by Gorkov and Melik-Bakhudarov is given by
\begin{equation}
\Delta/E_F= \left(\frac{2}{e}\right)^{7/3}e^{\pi/(2c)}.\label{eq:gapgorkov}
\end{equation}
In our approach we can extend this calculation to the strongly interacting regime and to the BEC side of the crossover. Our result for the gap in units of the Fermi energy, $\Delta/E_F$, is shown in Fig.\ \ref{fig:gap}. We find perfect agreement with Eq.\ \eqref{eq:gapgorkov} on the BCS side. At the unitarity point, $(ak_F)^{-1}=0$, we obtain $\Delta/E_F=0.46$. Another benchmark for the comparison of different methods is the chemical potential in units of the Fermi energy at $(ak_F)^{-1}=0$ and $T=0$. We find $\mu/E_F=0.51$ and compare the results of different approaches to this quantity as well as $\Delta/E_F$ in Tab. \ref{tab:fpvaluesSSB}.
\begin{table}[b!]
\caption{\label{tab:fpvaluesSSB}Results for the single-particle gap and the chemical potential at $T=0$ and at the unitarity point by various authors.}
\begin{ruledtabular}
\begin{tabular}{ccc}
&$\mu/E_F$ & $\Delta/E_F$ \\ \hline
Carlson \emph{et al.} \cite{MonteCarlo}& 0.43 & 0.54 \\
Perali \emph{et al.} \cite{Perali:2004zz}& 0.46 & 0.53\\
Haussmann \emph{et al.} \cite{Haussmann:2007zz}& 0.36 & 0.46 \\
Diehl \emph{et al.} \cite{Diehl:2007th}& 0.55 & 0.60 \\
Bartosch \emph{et al.} \cite{Bartosch:2009zr}& 0.32 & 0.61 \\
\emph{present work} & 0.51 & 0.46 \\
\end{tabular}\end{ruledtabular}
\end{table}

\subsection{Critical temperature}\label{crittemp}

\begin{figure}[b!]
\centering
\includegraphics[scale=0.99]{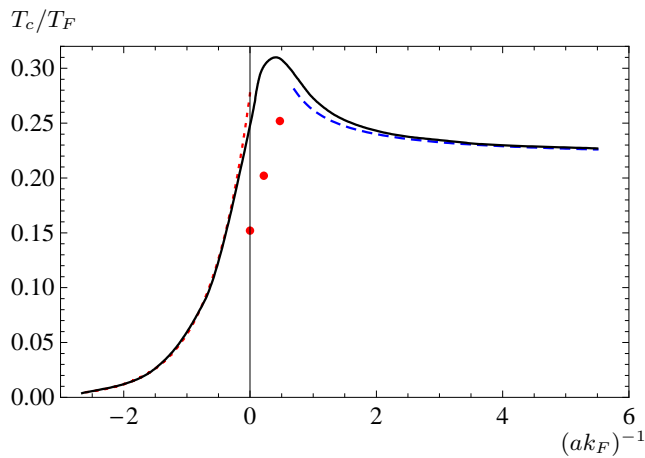}
\caption{Critical temperature $T_c/T_F$ in units of the Fermi temperature as a function of the crossover parameter $(ak_F)^{-1}$ (solid line). For comparison, we plot the BCS result with Gorkov's correction (left short dashed). On the BEC side we show the critical temperature for a gas of interacting bosonic molecules according to \cite{Baymetal,Arnold:2001mu,kashurnikov:2001} (dashed line on the right). The three red dots close to and at unitarity show the QMC results by  Burovski \emph{et al.} \cite{MonteCarloBuro}.}
\label{fig:tcrit}
\end{figure}
We next turn to the result for the critical temperature throughout the whole crossover which is shown in Fig.\ \ref{fig:tcrit}. On the BCS side ($c^{-1}<0$) we observe a perfect matching with BCS theory including the correction by Gorkov and Melik-Barkhudarov \cite{Gorkov}
\begin{equation}
\frac{T_c}{T_F}= 0.28 e^{\pi/(2c)},
\end{equation}
which is given by the short dashed line in Fig.\ \ref{fig:tcrit}. Deviations from this pertubative regime appear only quite close to the regime of strong interactions $c^{-1}\rightarrow 0$.

On the BEC-side our result approaches the critical temperature of a free Bose gas. This value is approached in the form \cite{Baymetal} 
\begin{equation}
\frac{T_c-T_{c,\mathrm{BEC}}}{T_{c,\mathrm{BEC}}}=\kappa\frac{a_M}{a}\frac{c}{(6\pi^2)^{1/3}},
\end{equation}
as shown by the long dashed line in Fig. \ref{fig:tcrit}. Here, $a_M$ is the scattering length between the molecules. For the ratio $a_M /a$ we use our result $a_M /a = 0.59$ obtained from solving the flow equations in vacuum. This has to be compared to the result obtained from solving the corresponding Schroedinger equation which gives $a_M /a = 0.6$ \cite{Petrov04}. For the coeffcient $\kappa$ determining the shift in $T_c$ compared to the free Bose gas we find $\kappa = 1.39$. In \cite{Arnold:2001mu,kashurnikov:2001} the result for an interacting BEC is given with $\kappa = 1.31$, which is in reasonable agreement with our result. Other characteristic quantities are the maximum of the ratio $(T_c/T_F)_\text{max}\approx0.31$ and the location of the maximum $(ak_F)^{-1}_\text{max}\approx0.40$.

In the unitary regime ($c^{-1}\approx 0$) our extensions of the truncation have a considerable quantitative effect. We can give an improved estimate for the critical temperature at the resonance ($c^{-1}=0$) where we find $T_c/T_F=0.248$
and a chemical potential $\mu_c/T_F=0.55$.
This compares to the previous FRG estimate $T_c/T_F=0.264$ and $\mu_c/T_F=0.67$ in ref.\ \cite{FSDW08}. Results from Quantum Monte Carlo simulations are $T_c/T_F = 0.15$ and $\mu_c/T_F=0.49$ in \cite{MonteCarloBuro}, $T_c/T_F < 0.15$ and $\mu_c/T_F=0.43$ in \cite{MonteCarloBulgac2} and $T_c/T_F = 0.245$ in \cite{MonteCarloAkki}. 

\section{Conclusions}
\label{sec:conc}

We have studied the flow equation approach to the BCS-BEC crossover in a rather detailed truncation. In addition to our previous treatments we have included contributions to the fermionic self-energy, a further interaction between fermions and bosons and improved our treatment of particle-hole fluctuations. Our results are in very good agreement with analytical treatments in various regimes. 

For small negative values of the concentration parameter $a k_F$ our findings for the gap at vanishing temperature $\Delta/E_F$ and the critical temperature $T_c/T_F$ confirm the predictions of Gorkov and Melik-Barkhudarov \cite{Gorkov}. On the other side, in the BEC regime with small positive values of $a k_F$, our results are in agreement with the expectations for a weakly interacting Bose gas where the shift in the critical temperature is linear in the dimer-dimer scattering length $a_M$. Also our finding for the ratio $a_M/a$ is in good agreement with the quantum-mechanical treatment \cite{Petrov04}. This quantitative accuracy is remarkable in view of the fact that we have started with a purely fermionic microscopic theory, without propagating bosonic degrees of freedom or bosonic interactions.

We present results for the single particle gap and the modifications of the Fermi surface at $T=0$ over the whole range of the crossover. This allows us to compute the dispersion relation for the single particle or fermionic excitations.
In the strongly interacting regime where the scattering length diverges, no analytical treatments are available. Our results for the gap $\Delta/E_F$ and the chemical potential $\mu/E_F$ at zero temperature are in reasonable agreement with Monte-Carlo simulations. This holds also for the ratio $\mu_c/E_F$ at the critical temperature. The critical temperature $T_c/T_F$ itself is found to be larger than the widely accepted Monte-Carlo result, however.

In future studies our approximations might be improved mainly at two points. One is the frequency- and momentum dependence of the boson propagator. In the strongly interacting regime this could be rather involved, developing structures beyond our current approximation. A more detailed resolution might lead to modifications in the contributions from bosonic fluctuations to various flow equations. Another point concerns structures in the fermion-fermion interaction that go beyond a diatom bound-state exchange process. Close to the unitarity point, other contributions might arise, for example in form of a ferromagnetic channel. While further quantitative modifications in the unitarity regime are conceivable, the present status of approximations already allows for a coherent description of the BCS-BEC crossover for all values of the scattering length, temperature and density by one simple method and microscopic model. This includes the critical behaviour of a second order phase transition as well as the temperature zero range.

\acknowledgments

We thank H.\ Gies and J.\ Braun for discussions and helpful comments. This work was supported by the DFG under FOR 723, and GK1523/1.

\end{document}